\documentclass[a4paper,aps,pre,twocolumn,groupedaddress,showkeys,showpacs,
  floats,floatats,floatfix]{revtex4-1}
\usepackage[utf8]{inputenc}
\usepackage[english]{babel}
\usepackage{graphicx}
\usepackage{dcolumn} 
\usepackage{bm}
\usepackage{natbib}
\usepackage{latexsym}
\usepackage{mathrsfs}
\usepackage{amssymb}
\usepackage{amsmath}
\usepackage{amscd}
\usepackage{color}
\usepackage{pifont}
\usepackage{ulem}
\usepackage{pstricks,pst-node,pst-text,pst-3d}
\usepackage{verbatim}
\usepackage[T1]{fontenc}

\newcommand{\RC}{{CRC}}
\newcommand{\QRC}{{QRC}}
\newcommand{\SRC}{{SRC}}

\newrgbcolor{Red}{1.0 0.0 1.0}
\begin{document}
\title{Quantum-classical transition and {quantum activation} of 
ratchet currents in the parameter space}
\author{M.~W.~Beims$^{1,2}$, M.~Schlesinger$^{3}$, C.~Manchein$^{4}$, A.~Celestino$^{2}$, A.~Pernice$^3$ and  W.~T.~Strunz$^3$}
\affiliation{$^1$Departamento de F\'\i sica, Universidade Federal do Paran\'a,
         81531-980 Curitiba,  Brazil}
\affiliation{$^2$Max-Planck-Institute for the Physics of Complex Systems, 
  N\"othnitzer Str.~38, 01187, Dresden, Germany}
\affiliation{$^3$Institute for Theoretical Physics, Technische Universit\"at Dresden, 01062 Dresden, Germany}
\affiliation{$^4$Departamento de F\'\i sica, Universidade do Estado de
  Santa Catarina, 89219-710 Joinville, Brazil} 
\date{\today}
%
\begin{abstract}
The quantum ratchet current is studied in the parameter space of the 
dissipative kicked {rotor} model coupled to a zero temperature
{quantum environment}. We show that vacuum fluctuations blur 
the generic isoperiodic stable structures found in the classical case.
Such structures tend to 
survive when {a measure of statistical dependence} between the 
quantum and classical currents are displayed in the parameter space. 
In addition, we {show that quantum fluctuations can be used to 
overcome transport barriers in the phase space. Related} quantum ratchet 
current activation {regions are spotted} in the parameter space. 
Results are discussed {based on quantum, semiclassical and classical
calculations. While the semiclassical dynamics involves vacuum fluctuations,
the classical map is driven by thermal noise.}
\end{abstract}
\maketitle

\section{Introduction}
\label{sec:introduction}

A general description of unbiased transport of particles in nature, which 
is named ratchet effect, is a challenging issue with implications in 
distinct areas, such as in molecular motors in biology \cite{julicher97}, 
nanosystems like graphene \cite{drexler13},  control of cancer metastasis 
\cite{campbell09}, micro and nanofluids \cite{boom11}, particles in silicon 
membrane pores \cite{matthias03}, cold atoms \cite{carlo05},  solids and 
drops transport using the Leidenfrost effect \cite{lagubeau11,Stout}, 
quantum systems \cite{grifoni97, dittrich01, klaus08,monteiro03}, among many others. 
Certainly the most relevant and common goal in describing these ratchet 
systems is to unveil how to control and attain an efficient transport 
by choosing the appropriate physical parameter combination, like temperature,
dissipation, external forces etc. {It is known \cite{dittrich01} that} 
in classical conservative systems the parameters must be chosen so that the 
underlying dynamics presents a mixture of regular and chaotic motion. In 
dissipative inertial systems the Classical Ratchet Current
(\RC) is more efficient when parameters are chosen inside the Isoperiodic 
Stable Structures (ISSs), which appear in the parameter space of ratchet 
models \cite{alan11-1,alan11-2}. Such ISSs {are generic Lyapunov stable 
islands with dynamics globally structurally stable and come along} 
in many dynamical systems \cite{jasonPRL93,kurths06,lorenz08,stoop10,gallas10}.
When stochastic effects are included, {\it e.~g}.~thermal fluctuations, these 
ISSs start to be destroyed and become blurred, even though they remain 
resistant to reasonable noise intensities \cite{alan13-1}. This means that in
realistic systems the \RC\,{is more efficient when parameters are 
chosen} inside the ISSs. In \cite{alan13-1} it was also shown that in certain 
multistability scenarios the current may actually be thermally activated.

The natural question now is whether the same general statements
can be made regarding the relation between the system's parameters (or the ISSs)
and the Quantum Ratchet Current (\QRC). With the ongoing 
technological 
developments engineering smaller and smaller devices {this question
becomes crucial for the observation of directed transport in quantum systems.}
It has been shown \cite{carlo12} that for specific points in the parameter 
space the \QRC\, apparently has the same main properties of the \RC\, inside 
the ISSs {when considered with a certain finite temperature} (see 
also \cite{carlo05}).
In this context one expects the quantum version of the ISSs to become blurred 
and gradually disappear as the full quantum limit is reached. In principle 
though, the simple addition of thermal effects in the classical dynamics is 
certainly not enough to reproduce the quantum dynamics. 

In the present work we calculate the \QRC\, using {a dissipative
zero temperature master equation. Therefore, fluctuations are of quantum 
origin while no thermal noise is considered in the description. In the 
semiclassical limit, the master equation can be recast as a semiclassical
map, allowing us to study in details the quantum-classical transition.
The semiclassical results are} compared to the classical map for the ratchet 
system, {derived through direct integration of the Langevin equation 
over a kicking period. The classical map involves correlated thermal
fluctuations.} We show how and to which extent the quantum-classical 
transition affects the optimal currents inside the ISSs. Even though vacuum 
fluctuations destroy the well defined shape of the ISSs when the current is 
plotted, the {borders of the} structures tend to survive when the 
quantum and classical currents {are statistically compared}~\cite{barikov07}.
In addition, the quantum fluctuations can assist current 
activation and we spot the activated regions in the parameter space.

{The paper is structured in the following way. In Sec.~\ref{sec:models}
we present the quantum, semiclassical and classical models used to investigate 
the ratchet current. In Sec.~\ref{sec:results} the results for the 
{quantum} ratchet current, {its activation,} and the distance 
correlations are shown in the parameter space and discussed. Section 
\ref{sec:conclusions} presents 
the summary of our findings and motivates related experiments.}

\section{Models}
\label{sec:models}

\subsection{The Quantum Problem} 

{The quantum problem}
can be described using the kicked Hamiltonian {in dimensionless
units}

\begin{equation}
\hspace{-0.05truecm}
\hat{\cal H}=\frac{\hat p^2}{2}+K\hspace{-0.1truecm}\left[\cos{(\hat x)}+
\tfrac{1}{2}\cos{(2\,\hat x+\tfrac{\pi}{2})} \right]
\sum_{m=0}^{+\infty}\hspace{-0.2truecm}\delta(t-m\tau),
\label{Qscaled}
\end{equation}
where ($\hat x, \hat p$) are position and momentum operators, 
$m=1,2,\ldots,N$ represents the discrete times when the ``kicks''
occur, $\tau$ is the kicking period, and $K$  is the parameter which 
controls the 
intensity of the kick. Dissipation is introduced between the kicks by 
coupling the ratchet system to a zero temperature  environment. 
In usual Born-Markov approximation {a master equation for the
reduced dynamics is obtained \cite{graham90}. Here} we determine the 
corresponding density operator as an ensemble mean over pure states 
obtained from the {corresponding} {\it quantum state diffusion} 
\cite{Gisin_Percival92} Ito-stochastic Schr\"odinger equation
\begin{eqnarray}
|d\psi\rangle & = & -{i}\hat{\cal H}|\psi\rangle dt 
+\sum_\mu\left(L_\mu - \langle L_\mu\rangle \right)
|\psi\rangle d\xi_\mu
\cr
& & \label{SE}\\
& & 
- \frac{1}{2}\sum_\mu
\left(L_\mu^\dagger L_\mu - 2\langle L_\mu^\dagger\rangle L_\mu + 
|\langle L_\mu\rangle|^2\right)|\psi\rangle dt,
\nonumber
\end{eqnarray}
where $\langle.\rangle$ stands for the expectation value.
The Lindblad operators {are given by}
$\hat L_1 = g\sum_n\sqrt{n+1}\left|n\right>\left< n+1 \right|,$
and $\hat L_2 = g\sum_n\sqrt{n+1}\left|-n\right>\left<-n-1 \right|,$
with  $g$ being the coupling constant and $n=0,1,\ldots$. The $|n\rangle$ 
states are the eigenstates of the momentum operator 
$\hat p|n\rangle = 2\pi\,n|n\rangle$. The Lindblad operators induce
a damping $-\lambda\langle \hat p\rangle$, with rate $\lambda=g^2$. 
{The semiclassical limit of
(\ref{Qscaled}) and (\ref{SE}) is obtained by taking $\tau\to0$ and
$\lambda\to\infty$, such that} the dissipation parameter 
$\gamma=e^{-\lambda\tau}$ {remains constant. It turns out that
 $\hbar_{eff}=\tau$ plays the role of the effective Planck constant 
\cite{carlo05}}. The \QRC\, is obtained from 
$\bf C_Q=||\langle\hat p\rangle||$,  where $||.||$ is a double average, over 
quantum realizations and time. As initial condition (IC) we use a coherent 
wave packet localized inside a minimum of the ratchet potential, 
{\it i.~e}.~$\left<\hat x_0\right>=\pi/2$ with $\langle\hat n_0 \rangle=0.0$. 

\subsection{The Semiclassical Problem}

We can better understand the quantum-classical relation regarding the 
directed transport by studying the \QRC\, as we approach the 
classical limit $\hbar_{eff}\to 0$. It is not reasonable, however, to
perform a full quantum calculation for this purpose because it is
numerically too demanding since more and more states and a huge number 
of time steps must be taken into account {in this limit}. To 
overcome this difficulty we use the semiclassical map proposed years 
ago \cite{graham90} for the standard map. In our ratchet case the 
semiclassical map at zero temperature takes the form
\begin{eqnarray} 
\tilde p_{m+1}&=&\gamma\,\tilde p_m + 
      \gamma\,K\left\{\sin{(\tilde q_m+\tilde\psi_{m+1})} \right.\cr
& &\left. \right. \cr
&+& \left.\frac{1}{2}\,\sin{\left[2(\tilde q_m+\tilde\psi_{m+1})
+\frac{\pi}{2}\right]}\right\} +\tilde\eta_{m+1}, \label{Smap}\\
\tilde q_{m+1}&=&\tilde q_m + 
\frac{(1-\gamma)\tau}{\gamma|ln\gamma|}\tilde p_{m+1}+\tilde\psi_{m+1},
\nonumber
\end{eqnarray}
where ($\tilde x_m,\tilde p_m$) are {scalars being the} position and 
momentum at discrete times $m=1,2,\ldots,N$, and $K$ 
{is the kicking parameter}. The map (\ref{Smap}) and the noise 
sources ($\tilde\psi_{m+1},\tilde\eta_{m+1}$) arise from conveniently writing the 
propagator of the Wigner function as a quasi-stochastic map (for more details 
see \cite{graham90}). The noise terms {are uncorrelated in time and} 
satisfy 
$\langle \tilde\eta_{m+1}\rangle=\langle\tilde\psi_{m+1}\rangle=0$, and 
\begin{eqnarray}
& &\langle \tilde\eta_{m+1}^2\rangle=\gamma\,(1-\gamma)\,\hbar_{eff}|\tilde p_m|/(2\pi),\cr
& & \cr
& &\langle \tilde\psi_{m+1}^2\rangle=\frac{4\,(1-\gamma)\,\hbar_{eff}}{\gamma|\tilde p_m|}
 +\frac{\gamma\,\hbar_{eff}|\tilde p_m|}{2\pi (1-\gamma)|ln\gamma|^2} 
 \left[|ln\gamma|^2 \right.\cr
& & \left. \right.\cr
& & \qquad \qquad\left. -2(1-\gamma) |ln\gamma|
 +(1-\gamma)^2(1-\frac{2}{\gamma}) |ln\gamma| \right.\cr
& & \left. \right.\cr
& & \qquad \qquad \left.+\frac{2}{\gamma}(1-\gamma)^3
 +\frac{(1-\gamma)^4}{\gamma^2}) \right],\cr
& & \cr
& &\langle \tilde\eta_{m+1}\tilde\psi_{m+1}\rangle=\frac{\gamma\,|\tilde p_m|\hbar_{eff}}{2\pi}
    \left[1-\frac{(1-\gamma)}{|ln\gamma|}-\frac{(1-\gamma)^2}{\gamma|ln\gamma|}
    \right].
\nonumber
\end{eqnarray}
Higher order cumulants were neglected {(higher orders in $\hbar_{eff}$)} 
so that the map (\ref{Smap}) can be interpreted as a classical map with 
Gaussian noise terms of quantum mechanical origin. This map is valid as long 
$\hbar_{eff}\ll1$. The current obtained from (\ref{Smap}) is referred to as 
\SRC\,(Semiclassical Ratchet Current). {ICs with 
$\langle\langle x_0\rangle\rangle = \langle\langle p_0\rangle\rangle = 0$ 
were taken inside the unit cell $(-2\pi,2\pi)$.} 

\subsection{The Classical Problem} 

The connection between the dynamics of a quantum system at zero temperature 
and its finite temperature classical limit in the parameter space
is not obvious at first sight.  Although one could argue that the 
first-order (in $\hbar$) quantum correction to a classical system is 
equivalent to a white noise term \cite{strunz1998classical}, the 
fluctuation-dissipation 
relation can differ, and it is also not clear whether a certain region of the 
parameter space would require higher order corrections for a reliable 
description of the system's dynamics. We access this connection by means of 
a direct comparison between the calculated current in parameter space for 
the quantum system and its classical finite temperature limit. This limit 
is obtained via integration of the {kicked Langevin} equation
\begin{eqnarray}
\dot p &=&-\lambda p+K\left [\sin{(x)}+\frac{1}{2}\sin(2\,x+\frac{\pi}{2})
\right ]\sum_{n=0}^\infty \delta(t-n\tau_c)\cr
& &+\sigma\xi(t),\cr
\dot x &=& p,
\nonumber
\end{eqnarray}
where $\tau_c$ is the classical kicking time, $\sigma = \sqrt{2\lambda k_B T}$,
and $\xi(t)$ is the white noise  satisfying $\langle \xi(t)\rangle=0$ and 
{$\langle \xi(t)\xi(s)\rangle=\delta(t-s)$.} The integration is performed 
between two kicks of the potential including the kick in the beginning, but 
not the one at the end. More precisely, the integration is performed between 
$t_0 = m - \epsilon$ and $t = m+1-\epsilon$, with positive 
$\epsilon\rightarrow 0$. Defining $\gamma = \exp(-\lambda\tau_c)$, 
the integration produces the following map 

\begin{eqnarray}
p_{m+1}&=&\gamma p_m+\gamma K\left[\sin{(x_m)}+\frac{1}{2}
\sin(2\,x_m+\frac{\pi}{2}) \right]+\alpha_m,\cr
& & \label{mapC}\\
x_{m+1}&=&x_m+\frac{(1-\gamma)\tau_c}{\gamma|\ln{\gamma}|}p_{m+1}+\beta_m,\cr
\nonumber
\end{eqnarray}
with 
the two stochastic terms ($\alpha_m,\beta_m$) featuring the 
properties $\langle\alpha_m\rangle =\langle\beta_m\rangle=0,$ and
\begin{eqnarray}
\langle\alpha_m^2\rangle &=& \frac{\,T_{eff}}{\tau_c}
(1-\gamma^2),\cr
\langle\beta_m^2\rangle &=&\frac{T_{eff}}{\gamma^2|\ln{\gamma}|^2}\left (2\gamma^2|\ln{\gamma}|+3\gamma^2-4\gamma+1 \right ),\cr
\langle\alpha_m\beta_m\rangle &=&-\frac{T_{eff}}{\tau_c\gamma|\ln{\gamma}|}(1-\gamma)^2,
\nonumber
\end{eqnarray}
with  $T_{eff}=k_BT\tau_c^2$.
In the map, the variables $x_m$ and $p_m$ represent respectively the position 
and the kinetic momentum of the particle when it is kicked by the $m$-th time 
($m=1,2,...,N$). Notice that the correlated thermal noise in the map 
(\ref{mapC}), {derived from the kicked Langevin equation with
white noise,} differs from the uncorrelated noise used in previous works 
\cite{carlo12,alan13-1}. As in the semiclassical case, ICs with 
$\langle\langle x_0\rangle\rangle = \langle\langle p_0\rangle\rangle = 0$ 
were taken inside the unit cell $(-2\pi,2\pi)$. 

\section{Results}
\label{sec:results}

\subsection{Ratchet current}

\begin{figure}[htb]
  \centering
 \includegraphics*[width=1.0\columnwidth]{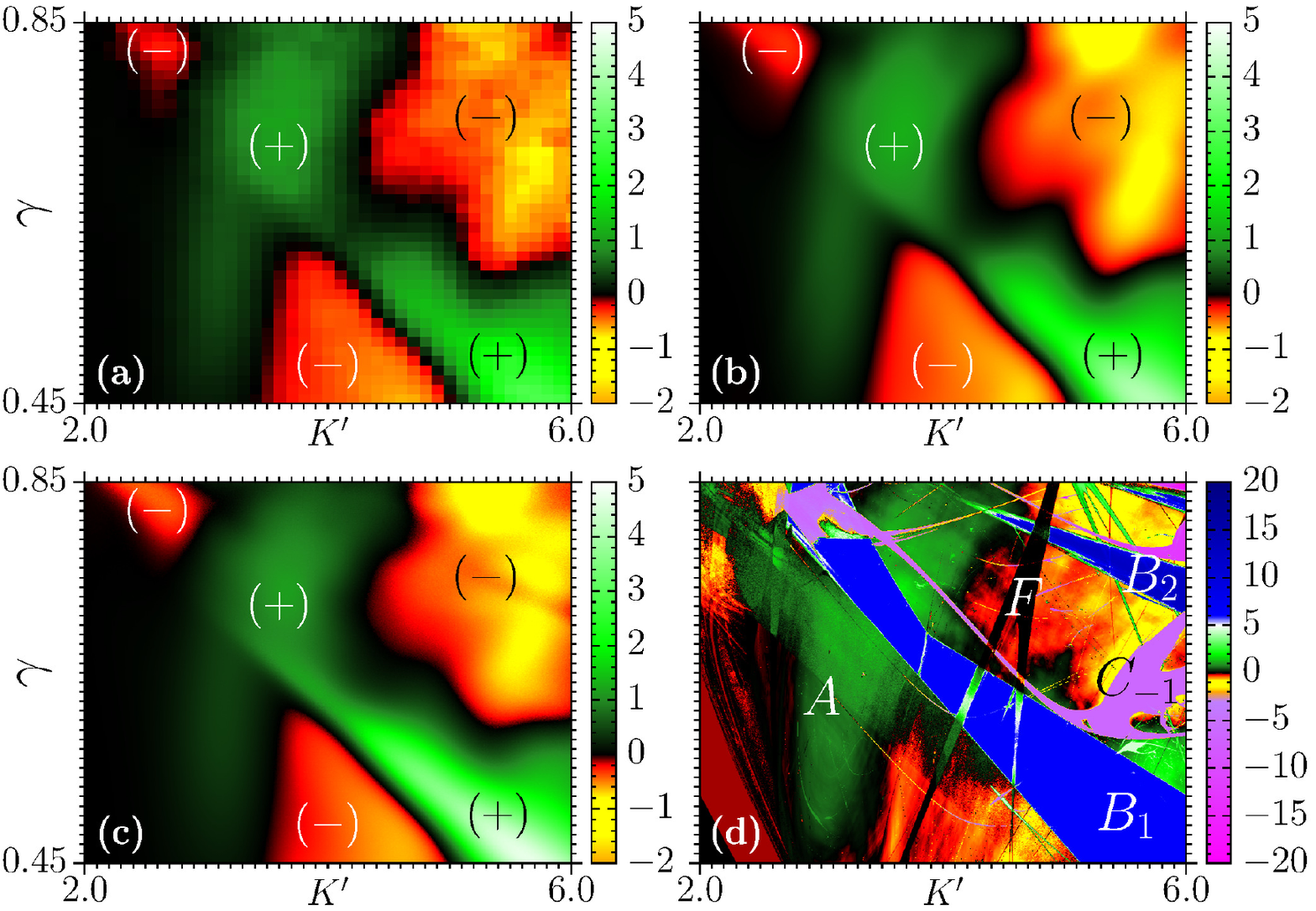}
 \caption{(Color online) The parameter spaces showing the (a) $\bf C_Q$,
(b) $\bf C_S$, for  $\hbar_{eff}=0.082$, and (c) $\bf C_C$\, for $T_{eff}=0.082$. 
Red (green) to yellow (white) colors are related to increasing negative 
(positive) currents [{marked with ($-$) and ($+$) for the printed 
grayscale]. Note the black boundaries delimiting positive and negative 
currents.}
In (d) we have the $\bf C_C$\, for $T_{eff}=0$ with black colors related to 
close to zero currents; green to blue colors (light to dark gray) are 
related to increasing positive currents while red, yellow to purple 
(white to light gray) colors related to increasing negative currents. 
Letters denote the chaotic background $A$ and the ISSs $B_1,B_2,C_{-1}$ 
and $F$.}
\label{PSS1}
\end{figure}

Now we can discuss results by comparing the currents of the three problems
above. Figure \ref{PSS1} shows the ratchet currents (see colors) (a) 
$\bf C_Q$ (quantum),\,(b) $\bf C_S$ (semiclassical)\,and (c)-(d) $\bf C_C$
(classical)\,in the parameter space
($K^{\prime},\gamma$) with $K^{\prime}=(1-\gamma)\tau\,K/|\ln{\gamma}|$
being the rescaled kick parameter. {In order to compare $\bf C_S$\, and
$\bf C_C$\, with the $\bf C_Q$,\, we 
used  ${\bf C_S}=(1-\gamma)\tau/(\gamma|\ln\gamma|)||\tilde p||$ and 
${\bf C_C}=(1-\gamma)\tau/(\gamma|\ln\gamma|) ||p||$, where the double 
average is over ICs and time. We checked that the $\bf C_Q$\, converges,
for the whole considered parameter space, after $10^3$ kicks and some
tens of stochastic realizations.} For the semiclassical and classical 
simulations we used $10^4$ iterations and $10^3$ ICs with zero average in 
position and momentum. It is worth to mention that for the parameter 
$\gamma$, shown in Fig.~\ref{PSS1}(a), the damping $\lambda$ is approximately 
inside the interval {[$2.0,9.7$]}, which represents strong dissipation. 
Thus the quantum dynamics is incoherent after 
the very short classical times $m_r\approx 1/(1-\gamma)$ \cite{graham90} 
and coherence effects are not expected to be relevant for the results
shown in the parameter space from Fig.~\ref{PSS1}(a).
Agreement between $\bf C_Q$\,[Fig.~\ref{PSS1}(a)] and $\bf C_S$\,[Fig.~\ref{PSS1}(b)]
is astonishing, meaning that the semiclassical map (\ref{Smap}) nicely 
reproduces the quantum results in the quantum-classical transition for 
this range of parameters. Note that the $\bf C_C$\, in Fig.~\ref{PSS1}(c), 
obtained using one fixed $T_{eff}$ in the classical map (\ref{mapC}),
also shows a qualitative agreement with the $\bf C_Q$.

To understand the physical origin of the distinct currents in 
Figs.~\ref{PSS1}(a)-(c) we analyze the $\bf C_C$\, at zero temperature $T_{eff}=0$. 
This is shown in  Fig.~\ref{PSS1}(d).
Two main regions with distinct behaviors can be observed: first, the 
{\it ``cloudy''} background, identified as $A$ in Fig.~\ref{PSS1}(d) showing 
a mixture of zero, small negative and positive currents, where the 
dynamics is chaotic \cite{alan11-2}. Second, the four ISSs, 
$B_1, B_2$, $C_{-1}$ and $F$. Besides $F$, which has zero 
current, the ISSs are responsible for the optimal $\bf C_C$\,and can be 
recognized by their sharp borders (for a detailed explanation see 
\cite{alan11-1,alan11-2}). We note that at the crossing regions
of $B_1$ and $F$, positive currents are observed as a consequence of
multistable attractors. The portion of the parameter space shown in 
Fig.~\ref{PSS1} can be 
regarded as representative of the highly dissipative ratchet regime of 
this system, presenting  the main properties found in the larger parameter 
space, and is thus suitable for the purpose of the present work. Comparing 
Figs.~\ref{PSS1}(a)-(c) to Fig.~\ref{PSS1}(d), we can identify that the 
positive and negative $\bf C_Q$, $\bf C_S$ and $\bf C_C$ ($T_{eff}>0$) roughly follow the 
overall chaotic currents from the region $A$, but are enhanced inside the 
ISS $B_1$ from the classical case with $T_{eff}=0$. This shows that vacuum 
fluctuations tend to blur the classical ISSs, as suggested recently 
\cite{carlo12}. 
\begin{figure}[htb]
  \centering
 \includegraphics*[width=1.0\columnwidth]{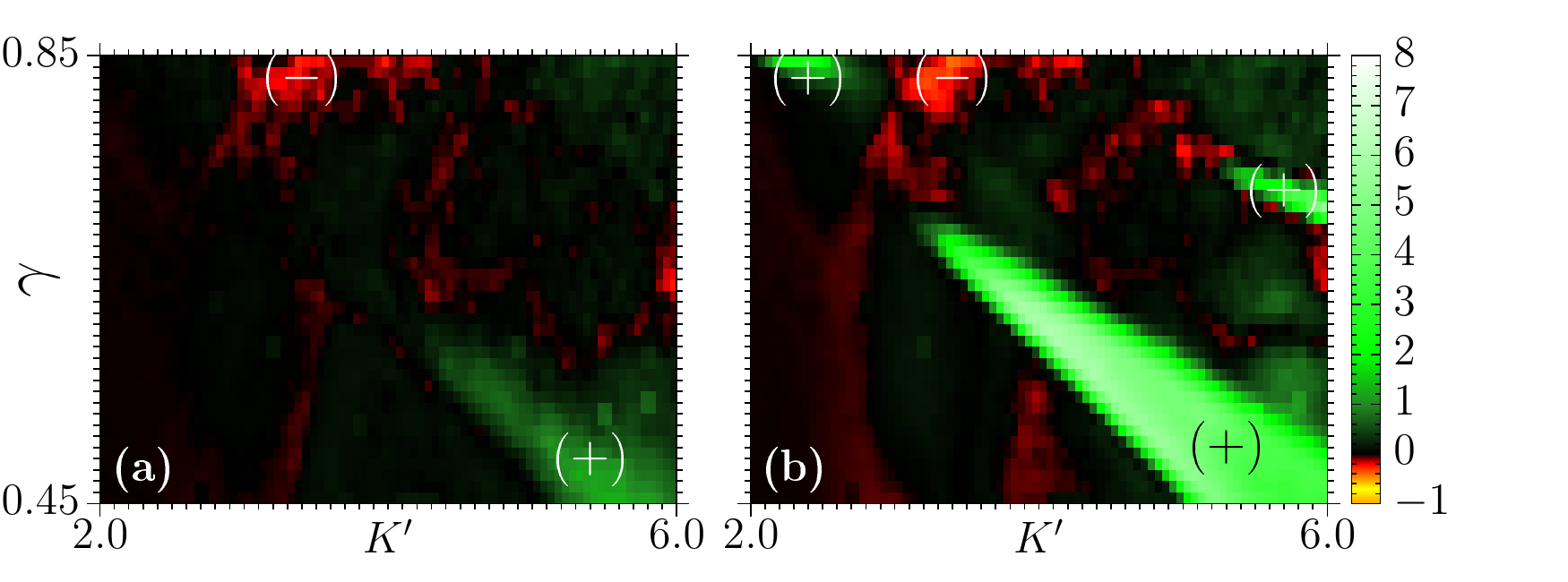}
 \caption{(Color online) Plotted is (a) $|\mbox{$\bf C_S$}|-|\mbox{$\bf C_Q$}|$
 and (b) $|\mbox{$\bf C_C$}|-|\mbox{$\bf C_Q$}|$ from the data of Fig.~\ref{PSS1}
 [Symbols ($-$)/($+$) to identify the negative/positve currents in 
 printed grayscale].}
  \label{abc}
\end{figure}
{One can recover many of the $\bf C_Q$'s features in parameter space using
classical calculations. This was also observed in \cite{carlo12}, for specific 
points in the parameter space, in a system in contact with a distinct thermal 
bath with uncorrelated noise. However, while a purely classical map with 
thermal fluctuations (and single $T_{eff}$ value) leads to a $\bf C_C$\, that can 
qualitatively mimic $\bf C_Q$\, in parameter space, it is usually quantitatively 
very far off. Much better agreement, both qualitatively and quantitatively, 
is reached using a semiclassical map. This is shown in Fig.~\ref{abc}.
We stress that quantum results are very well
represented by the semiclassical map without any free parameter. By
contrast, the simulation in terms of the classical approach (\ref{mapC}),
involves $T_{eff}$ as an additional free parameter that needs to be adjusted
for every point in the parameter space to achieve an overall agreement
with the $\bf C_Q$. Thus}, it becomes clear that a single $T_{eff}$ is not able 
to reproduce the $\bf C_Q$\, for the whole parameter space.

\subsection{Correlation between \SRC\, and \RC}

A result for the quantum-classical transition regarding ratchet currents 
is found by analyzing the 
correlation between the $\bf C_S$\,(at $\hbar_{eff}\ne 0$)  and the $\bf C_C$\,(at 
$T_{eff}=0$). It is a measure of statistical dependence,
{called distance correlation \cite{barikov07}}, between both currents, 
which is zero if and only if the currents are statistically independent. 
Thus, the purpose of the present analysis is to search to which extent
the quantum current is statistically similar to the classical current.
For this we use the distance correlation, which is obtained from the expression 
\cite{barikov07} 
${\cal D}{(\tilde p,p)}= d{(\tilde p,p)}
/\sqrt{\Delta({\tilde p})\Delta({p})}$, 
where 
\begin{equation}
d{(\tilde p,p)}=\frac{1}{N}\sqrt{\sum_{j,k=1}^N\,{S}_{j,k}{C}_{j,k}}
\end{equation}
is the distance covariance, and 
\begin{equation}
\Delta(\tilde p) = \frac{1}{N}\sqrt{\sum_{j,k}^N\,({S}_{j,k})^2},\quad\Delta(p)=
\frac{1}{N}\sqrt{\sum_{j,k}^N\,({C}_{j,k})^2}, 
\end{equation}
are the distance variances. Here $\tilde p$ refers to the time sequence 
${\tilde p}_1, \tilde p_2,\tilde p_3,\ldots,\tilde p_m$ obtained from the 
map (\ref{Smap}), leading to the $\bf C_S$.  $p$ refers to the time sequence 
$p_1, p_2, p_3,\ldots, p_m$ 
obtained from the map (\ref{mapC}), leading to the $\bf C_C$. $S$
and $C$ are matrices defined through
${S}_{j,k}=s_{j,k}-\overline{s}_{j.}-\overline{s}_{.k}+\overline{s}_{..}$
and ${C}_{j,k}=c_{j,k}-\overline{c}_{j.}-\overline{c}_{.k}+\overline{c}_{..}$,
respectively. Here $s_{j,k}= |\overline{\tilde p}_j-\overline{\tilde p}_k|$ 
is the Euclidean norm of the distance between the momenta (averaged over ICs) 
at times $j$ and $k$. Thus, $\overline{\tilde p}_j$ is an  average over ICs 
at times $j$. The same is valid for $c_{j,k}= |\overline{p}_j-\overline{p}_k|$. 
In addition, $\overline{s}_{j.}$ is the $j$-th row mean, $\overline{s}_{.k}$ 
the $k$-th column mean, and $s_{..}$ is the mean value between 
$\overline{s}_{j.}$ and $\overline{s}_{.k}$ (the same notation applies to $c$). 
Thus, each element
of the matrices $S$ and $C$ contains information about: distances
{\it in time} of the currents (already averaged over ICs) and also time averages 
of these distances over the time series $\tilde p$ and $p$. The distance correlation 
${\cal D}{(\tilde p,p)}$, on the other hand, analyses the statistical independence 
of both matrices. It is not a direct quantity to be measured in one simulation (or 
experiment), but compares the currents obtained by two simulations performed in 
distinct semiclassical regimes.

\begin{figure}[htb]
  \centering
 \includegraphics*[width=1.0\columnwidth]{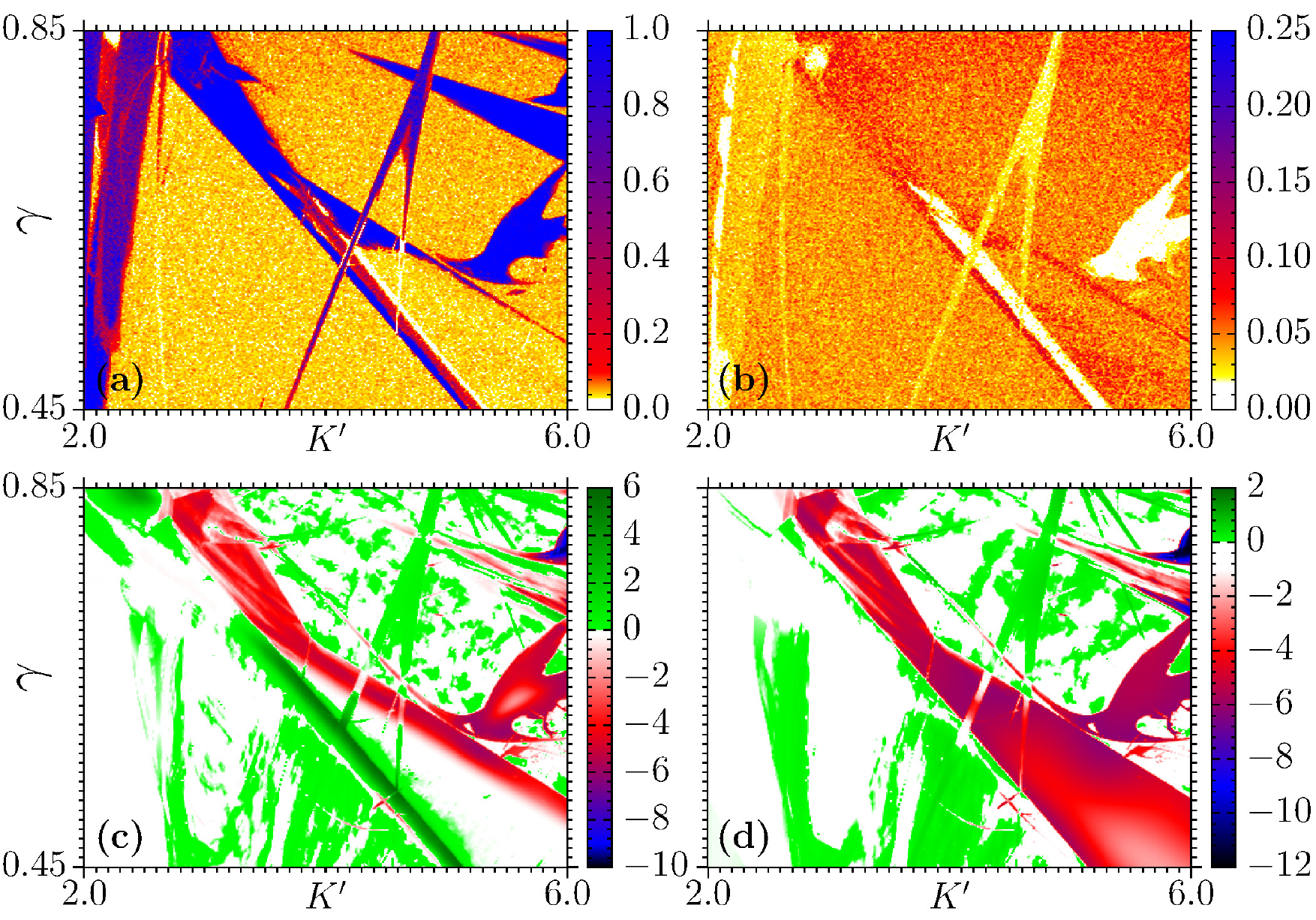}
 \caption{(Color online) Plotted is  ${\cal D}{(\tilde p,p)}$ 
for (a) $\hbar_{eff}=0.0001$ and (b) $\hbar_{eff}=0.082$, and the
activation/suppression $\delta(\hbar_{eff})$ for (c) $\hbar_{eff}=0.01$  
and (d) $\hbar_{eff}=0.082$.}
  \label{Dc}
\end{figure}

In Figs.~\ref{Dc} (a) and (b) the quantity ${\cal D}{(\tilde p,p)}$ is displayed 
(colors) in the same parameter space from Fig.~\ref{PSS1}. White, yellow, red to 
blue (white, light to dark-gray) colors are related  to increasing values of the 
distance correlation. For each point in the parameter 
space we used $N=4.5\times 10^3$ points of the time sequence of $\tilde p_m$ and 
$p_m$, averaged over $5\times 10^4$ ICs. 
 At small $\hbar_{eff}=0.0001$ we observe in the 
parameter space for ${\cal D}{(\tilde p,p)}$ that the 
structures $B_1,B_2,C_{-1}, F$ can be clearly recognized (they have sharp 
borders). For parameters inside the ISSs the quantum-classical dynamics is 
strongly correlated, while outside the ISSs the correlation goes to zero. 
Inside the ISS $B_1$ we observe that correlations also go to zero, but its
border is still well defined. For $\hbar_{eff}=0.082$ the result is inverted,
as inside most part of the ISSs the correlation is zero and outside it remains 
almost the same as from Fig.~\ref{Dc}~(a). In this case the ISS $B_2$ 
disappeared, the $C_{-1}$ and $F$ yet exist with sharp borders, and some 
remaining borders of $B_1$ are visible. These results demonstrate that the 
{\it sharp borders} of the ISSs tend to survive in the parameter space 
(for reasonable $\hbar_{eff}$) when ${\cal D}{(\tilde p,p)}$ is plotted, 
while the ISSs are already blurred and unrecognizable when the ratchet current
is plotted [compare Fig.~\ref{Dc}(b) to Figs.~\ref{PSS1}(a) and (b)]. 
In other words, the way the ISSs disappear in both cases is fundamentally
distinct. It is obvious that using huge values for $\hbar_{eff}$ the 
correlation in the whole parameter space will go to zero. 

\subsection{Quantum activation of the ratchet current}

Another interesting related phenomenon is that the quantum fluctuations 
not only lower the $\bf C_S$, but also activate them. In order to show this we 
calculate 
$\delta(\hbar_{eff})=|\mbox{$\bf C_S$}(\hbar_{eff})|-|\mbox{$\bf C_C$}(T_{eff}=0)|$ and 
plot a map of the current activation/suppression in the parameter space. 
For $\delta>0$ activation of the $\bf C_S$\, occurs while for $\delta<0$ the 
$\bf C_S$\, decreases due to vacuum fluctuations. Results are shown in 
Fig.~\ref{Dc}(c) and (d) for the same parameter space from  Fig.~\ref{PSS1}. 
We use green to dark-green (gray to dark-gray) colors for increasing activations
of the $\bf C_S$\,and white, red, blue to black (white, dark-gray to black) for 
increasing suppressions of $\bf C_S$.
Surprisingly there are many regions in the parameter space where activations
occur. The physical origin of the activations depend on the specific
dynamics at a given parameter combination. Along the diagonal dark-green
line (below the ISS $B_1$) where large activations ($\Delta\sim 6$) occur
in Fig.~\ref{Dc}(c), we conclude from classical results \cite{alan11-1} 
that the activation is due to a crisis bifurcation. Additional numerical 
results revealed that the activation close to $K=2.5$ and $\gamma=0.85$ 
is related to a symmetry breaking of the multistable attractors.

\section{Conclusions}
\label{sec:conclusions}

Concluding, the quantum ratchet current was analyzed in the parameter
space of the asymmetrically kicked rotor. We have shown that  
isoperiodic stable structures with sharp borders play an essential role 
in the quantum-classical transition. Quantum currents were obtained {
{numerically exact} by solving the Markov stochastic Schr\"odinger 
Eq.~(\ref{SE})($\bf C_Q$), and in a semiclassical approach from the 
semiclassical map (\ref{Smap})($\bf C_S$).} For the classical currents ($\bf C_C$) we 
used the map (\ref{mapC}) which 
was derived via direct integration of the kicked Langevin equation 
with a white noise. {The quantum-classical transition was studied
by means of a comparison between the $\bf C_S$\, and $\bf C_C$.}
We show how and to which extent the classical generic isoperiodic 
stable structures  \cite{alan11-1,alan11-2,alan13-1} are destroyed by quantum 
fluctuations. Even though these structures become blurred and start to 
disappear in the quantum world, their sharp borders tend to persist when 
the correlation between quantum and classical currents is investigated. 
{For this aim we have used the distance correlation measure 
\cite{barikov07}. Remarkably we have found a statistical dependence of the
quantum and classical currents when parameters are chosen at the borders
of the isoperiodic stable structures. Moreover, very often this dependence
extends to the interior of the isoperiodic structures.} In addition, we 
calculated the quantum activation/suppression of the current in the parameter 
space. Surprisingly many regions {were found where} quantum fluctuations 
{enhance the $\bf C_C$. Even though the distance correlation and the current activation 
are not directly observed for one simulation or experiment, they can be obtained
by comparing two simulations in distinct semiclassical regimes and allow
us to understand better the quantum-classical transition of ratchet currents.}

It is generally accepted that dissipative quantum dynamics inducing
decoherence will blur any quantumness and eventually will lead to a
dynamics that may well be described in classical terms. In this paper 
we show, however, that for the ratchet current in a zero-temperature
environment, this quantum-classical transition is far from trivial 
and depends crucially on the chosen parameters. We prove that a 
replacement of quantum fluctuations by classical thermal fluctuations 
is overly simplistic and cannot be used for the whole parameter space
uniformly. It is clear that a single quantity like the $\bf C_Q$\, can be 
matched by a $\bf C_C$\, if one allows an additional free parameter (here 
temperature) to vary. Crucially, however, different points in 
parameter space require very different temperatures for this fit 
procedure (see Fig.~2), and in this sense a quantum-classical 
correspondence, where quantum fluctuations are replaced by thermal 
fluctuations, cannot exist. In addition, there is no one scaling
which gives the quantum-classical correspondence in the whole 
parameter space. One of the reasons for this behavior can be traced 
back to the complicated dynamics in the case of mixed phase space 
structure. In our context we can write
$\bf C_S = \sum_i \alpha_i\bf C_S^{(i)}$, where the sum is over the attractors,
$\bf C_S^{(i)}$ is the ratchet current from attractor $i$ weighted by the 
area of the corresponding basin of attraction, and $\alpha_i=\mu_i\tau_i$ 
is the statistical weight of each attractor, where $\tau_i$ is the mean 
lifetime on attractor $i$ and $\mu_i=lim_{n\to\infty}\langle m_i\rangle/n$ 
where $m_i$ is the number of times a given trajectory visited the attractor
$i$ during the time $n$. Quantum fluctuations allow the trajectory to 
jump between attractors and $\alpha_i$ is the quantity which depends, 
in a nontrivial way, on $h_{eff}$ and may also depend on the size/shape
of the basin of attraction. The degree of quantum-classical correspondence 
is therefore strongly dependent on the local dynamics, and general 
statements about the scaling of $\alpha_i$ are difficult. However, our 
results indicate a qualitatively different correspondence for parameters 
inside and outside the ISS regions. Indeed, we have shown that the way the 
ratchet current responds to the quantum fluctuations depends sensitively 
on whether the parameters are chosen within the ISSs' limits. It indicates 
that the classical-quantum transition is much more abrupt inside the ISSs, 
at least regarding the ratchet current. 

{Our results are an important step towards the goal to experimentally
observe quantum isoperiodic structures} in the context of ratchet 
effects with cold atoms \cite{tabosa99,sandro12}. {Moreover, we
strongly believe that quantum activations deserve further analysis, both
theoretically and experimentally.} In general, the analysis 
of the {complex} quantum-classical transition in the parameter
space (including the isoperiodic structures), is not restricted to 
ratchet currents. Other observable of a quantum system,  whose classical 
counterpart presents regular and chaotic dynamics, could be used. 
We mention for example the motion of atoms in an optical lattice 
\cite{garreau05,garreau11} and the electronic wave-packet dynamics in
Rydberg atoms subjected to a strong static magnetic field 
\cite{alber,beims93,beims96}.

\section{Acknowledgments} 
Our international cooperation is supported by 
PROBRAL - DAAD/CAPES and M.W.B. and C.M. thank CNPq for financial support
and M.W.B. thanks MPIPKS in the framework of the Advanced Study Group on 
Optical Rare Events.

\vspace*{-0.8cm}

\end{document}